\documentclass[aps,prd,amsmath,amssymb,superscriptaddress,nofootinbib,10pt,oneside]{revtex4}
\usepackage[]{latexsym}
\usepackage[english]{babel}
\usepackage{graphicx}
\usepackage{hyperref}
\usepackage{color}
\usepackage{bm}

\newcommand{\Dk}[1]{\frac{d^3#1}{(2\pi)^3}}

\newcommand{\ve}[1]{{\text{\bf #1}}} 
\newcommand{\vk}{\ve k}
\newcommand{\vp}{\ve p}
\newcommand{\vq}{\ve q}
\newcommand{\vx}{\ve x}

\newcommand{\Ps}{\mathbf{\Psi}}

\begin{document}
%openinmog
% % \title{Renormalization of Lagrangian multivariate halo bias through spectral parameters with applications to generalized cosmologies}
\title{Renormalization of Lagrangian bias via spectral parameters}
\author{Alejandro Aviles}
\email{avilescervantes@gmail.com}
\affiliation{Departamento de F\'isica, Instituto Nacional de Investigaciones Nucleares,
Apartado Postal 18-1027, Col. Escand\'on, Ciudad de M\'exico,11801, M\'exico.}
\affiliation{Consejo Nacional de Ciencia y Tecnolog\'ia, Av. Insurgentes Sur 1582,
Colonia Cr\'edito Constructor, Del. Benito Jurez, 03940, Ciudad de M\'exico, M\'exico}
% \pacs{PACS}

\begin{abstract}

We extend the definition of Lagrangian local bias proposed by Matsubara (2008) to include curvature and higher-derivative bias operators. 
Evolution of initially biased tracers using perturbation theory (PT) generates multivariate bias parameters
as soon as nonlinear fluctuations become important. We present a procedure
that reparametrizes a set of spectral parameters, the arguments of the Fourier transformed Lagrangian bias function, 
from which multivariate renormalized biases can be derived at any order in bias expansion and PT. 
We find our method simpler than previous renormalization 
schemes because it only relies on the definition of bias, fixed from the beginning, and in 
one equation relating renormalized and unrenormalized spectral parameters. 
We also show that our multivariate
biases can be obtained within the peak background split framework, in that sense this work extends that of
Schmidt, Jeong and Desjacques (2013). However, we restrict our method to Gaussian initial conditions. Non-linear evolution also leads
to the appearance of products of correlators evaluated at the same point, commonly named contact terms, yielding
divergent contributions to the power spectrum. In this work we present an explicit method to remove these divergences
by introducing stochastic fields.

\end{abstract}

\maketitle

\begin{section}{Introduction}

Upcoming galaxy surveys such as DESI \cite{Aghamousa:2016zmz}, Euclid \cite{Amendola:2012ys} and WFIRST \cite{Spergel:2015sza}
will impact our understanding of the evolution of the Universe by measuring 
with high precision the cosmological parameters at low redshifts, and also they are likely to answer more fundamental questions such
as the value of the mass of neutrinos or even to test gravity at cosmological scales. As the depth and size of the surveys increases they cover
scales where quasi-linear effects are more relevant and the tools of perturbation theory (PT) become even more important. 
To fully exploit the already existing and forthcoming wealth of data within analytical and semianalytical methods, 
a concise theory of clustering is needed. With the exception of weak lensing, the dark matter clustering is not observable directly, but it should
be deduced from the clustering of galaxies, Ly-alpha forest, and other biased tracers of the 
underlying matter content \cite{Kaiser:1984sw,Desjacques:2016bnm}. PT
of matter fluctuations is well understood within its range of validity \cite{Bernardeau:2001qr}, but
this is not the case for the PT of tracers, which requires
the inclusion of information about halos and galaxy formation and evolution. Being these highly non-linear processes, they are 
apparently out of the reach of PT. However, biased tracers can be described within PT as an effective field theory (EFT)
with a set of unknown parameters (the bias parameters) that are in principle free and should be determined by observations or simulations.
The situation becomes more complicated since the bias parameters evolve in general with time and scale \cite{Fry:1996fg,Hui:2007zh}. 
An EFT smooths the relevant fields by removing out of the theory their small scales. Since the smoothing scale $R_\Lambda$ (or equivalently
$\Lambda=1/R_\Lambda$) is arbitrary, and hence unphysical, it should not appear in observables such as statistics of tracers; this reasoning led McDonald to 
propose a renormalization procedure of bias parameters \cite{McDonald:2006mx}. 
The first description of bias relied on locally expanding the overdensity of tracers
in powers of the matter overdensities $\delta_m$ \cite{Kaiser:1984sw,Fry:1992vr,Matarrese:1997sk,Dekel:1998eq}. 
Soon, several authors realized that this procedure had some theoretical flaws,
since for example quantities such as $[\delta_m]^2$ were not necessarily smaller than $\delta_m$; thus the introduction of nonlocal bias operators in the
bias expansion was required and the process of renormalization 
has been extended \cite{2009JCAP...08..020M,Matsubara:2011ck,Schmidt:2012ys,Assassi:2014fva}. 

The bias expansion can be performed on either evolved or initial density fields; the former is named Eulerian bias and the latter as 
Lagrangian bias. In the Lagrangian approach it is assumed that the initial overdensities are linear at all scales of interest, such that 
one can guarantee that $\delta_m(t_{ini}) \ll 1$, and a local expansion in matter densities is at least well defined; other contributions
such as tidal bias can be generated by subsequent nonlinear evolution \cite{Baldauf:2012hs}. This does not mean Lagrangian tidal bias or
other nonlinear biases should not be incorporated,
because nonlinearities, although negligibly small, are still present and will eventually dominate the clustering of matter. Moreover,
if we let tidal contributions be generated only by the gravitational evolution, they will appear in the evolved fields carrying the local bias 
parameters, while in principle they should carry their own bias parameters. 
Lagrangian tidal bias has been considered by some authors \cite{Castorina:2016tuc,Vlah:2016bcl,Abidi:2018eyd} and currently there
is good evidence that it is nonzero \cite{Modi:2016dah}. In this work we will not consider tidal bias, 
but we foresee no obstacles to introduce it following the
path of \cite{Vlah:2016bcl}. 

The main subject of this work is the local (in mass density) Lagrangian description and that, as has been noted by 
Schmidt, Jeong and Desjacques \cite{Schmidt:2012ys}, besides standard renormalization that removes zero-lag correlators, 
it needs the inclusion of curvature $\nabla^2 \delta$ and higher derivatives $\nabla^{2N}\delta$ 
in order to remove subleading dependencies on the smoothing scale $R_\Lambda$. Our approach assumes the existence of a 
Lagrangian bias function relating overdensities of matter and tracers, $1+\delta_X = F(\delta, \nabla^2 \delta, \nabla^4 \delta, \dots)$; each argument generates
a set of \emph{univariate} bias parameters: $c_{n00\dots}$ for $\delta$, $c_{0m0\dots}$ for $\nabla^2\delta$, and so on. As long as
the evolution remains linear those are all the parameters we need, but when nonlinear fluctuations become important \emph{multivariate} biases
$c_{nm\dots}$ with both $m$ and $n$ different from zero should be included.\footnote{We notice that the name \emph{multivariate bias} 
and the notation $c_{nm}$ have been used
in different contexts \cite{Giannantonio:2009ak,Matsubara:2011ck,Schmidt:2012ys,Bardeen:1985tr,Desjacques:2010gz}, 
as in bias from non-Gaussianity and the peak model.}
Typically, the $c_{nm\dots}$ are derivatives of 
the function $F$ evaluated at zero values of the arguments. This description leads to the renormalization of the $c_{nm}$ biases, 
as much as it happens for the univariate bias \cite{Schmidt:2012ys}. 

In \cite{Matsubara:2008wx}, Matsubara put forward a closely related procedure for local bias parameters which takes as its most important object
the argument of the Fourier transformed local Lagrangian bias function $\tilde{F}(\lambda)$, 
that we name here the local spectral bias parameter $\lambda$. 
The bias local parameters at $n$ order in the bias expansion are obtained by simple integrations of powers of $\lambda$. 
It turns out that the local biases derived in this way 
are automatically renormalized in the sense that $N$-point statistics 
have no zero-lag correlators. In this work we generalize this procedure to multivariate biases; hence our principal objects of interest
are a set of spectral bias parameters $\lambda, \eta_{\nabla^2\delta}, \eta_{\nabla^4\delta}$, corresponding to the arguments of the Fourier transformed
nonlocal Lagrangian function.  Although describing bias in terms of ``space'' or spectral parameters is equivalent, 
we find the latter economically simpler; for example,
a relation between bare and renormalized local bias can be obtained in a single line [see Eq.~(\ref{cnTobn})].
However, we shall note that the multivariate biases obtained in this way need renormalization, unlike the $b_n=c_{n0\dots}$ obtained from the spectral
parameter $\lambda$ only. In this work we present a renormalization method that reparametrizes directly the spectral parameters, 
instead of the bias parameters themselves, with the advantage that it only needs one  relation 
[Eq.~(\ref{spectralbiasrel})] to renormalize any multivariate
bias parameter $c_{n_1n_2\cdots n_N}$. We further show that our renormalized bias parameters can be obtained within the framework of peak background 
split \cite{Kaiser:1984sw,Mo:1996cn,Sheth:1999mn}, where the bias parameters measure the changes 
of the mean abundance of tracers against small constant shifts in background density and in curvature \cite{Schmidt:2012ys}.

We use Lagrangian perturbation theory (LPT) \cite{Zel70,Buc89,Mou91,TayHam96} to evolve the initially biased tracers, and the resummations leading to 
standard perturbation theory (SPT) \cite{Matsubara:2007wj} and convolution Lagrangian perturbation theory (CLPT) \cite{Carlson:2012bu} to 
obtain the 1-loop SPT power spectrum and CLPT correlation function, respectively. 
Nonlinear evolution of fluctuations leads to the appearance of the product of correlators evaluated at the same point, commonly named contact 
terms following the usage in field theory. When Fourier transformed, these terms have UV divergences that is well known can be removed by
the introduction, and {\it a posteriori} renormalization of stochastic fields \cite{Dekel:1998eq,Taruya:1998hf,Matsubara:1999qq} 
and corresponding bias parameters \cite{2009JCAP...08..020M,Assassi:2014fva}. 
In this work we present a systematic procedure to remove the UV divergences from any
contact term by adding a finite collection of counterterms that are ``absorbed'' by the stochastic fields.

We organize this work as follows. In Sec. \ref{Sect:2} we present results for locally biased tracers and its 
nonlinear evolution within PT. Some of these results are known from the works of \cite{Matsubara:2008wx,Carlson:2012bu,Matsubara:2007wj}, 
but we give some insights in order to generalize them in the subsequent sections. We further present the renormalization of the first contact term, 
following Ref. \cite{McDonald:2006mx}. In Sec. \ref{Sect:3} we generalize the definition of bias to include curvature and higher order derivative operators,
thereafter we present our method of renormalization via spectral bias parameters. 
In Sec. \ref{Sect:4} we discuss the UV divergences in the power spectrum coming from 
Fourier transformed contact terms and we show how these can be removed by stochastic fields. 
We conclude in Sec. \ref{Sect:conclusions}.

% Even for the case of baryions some sort f biasing may be considered due to selection effects.

%\cite{Brax:2013fna}
% \cite{Aviles:2017aor} 

%\cite{Matsubara:2008wx}

\end{section}

\begin{section}{Locally biased tracers and their non-linear evolution}\label{Sect:2}

We consider particles with (Lagrangian) position $\vq$ at some early time $t_{ini}$; the (Eulerian) position
$\vx(\vq,t)$ at a later time $t$ is given by the transformation rule
\begin{equation}\label{coordtrans}
 \vx(\vq,t) = \vq + \Ps(\vq,t),
\end{equation}
where $\Ps$ is the Lagrangian displacement vector and $\Ps(q,t_{ini})=0$. 
Matter conservation allows us to write the fluid overdensities as \cite{TayHam96}
\begin{equation}\label{deltam}
 \delta_m(\vk) = \int d^3q e^{-i\vk \cdot \vq}\left( e^{-i\vk \cdot \Ps(\vq,t)} -1 \right), 
\end{equation}
as long as the initial overdensities are sufficiently small, $\delta(\vq) \ll 1$.
The transverse piece of the Lagrangian displacement is nonzero starting at third order in PT if velocity dispersions and higher momenta
can be neglected \cite{Matsubara:2015ipa,Aviles:2015osc,Cusin:2016zvu}. In this work we deal with 2-point statistics (up to 1-loop) 
of cold dark matter particles; hence we can treat $\Ps$ as longitudinal. We further assume that the linear displacement field is drawn from 
a Gaussian distribution. To linear order in fluctuations we get
\begin{equation}
 \Psi_{i,i}^{(1)}(\vq,t) = - \delta_{L}(\vq,t)
\end{equation}
where $\delta_{L}(\vq,t)$ is the linearly extrapolated initial matter overdensity  $\delta_{L}(\vq,t)= D_+ (t) \delta(\vq)$, with $D_+$ the 
linear growth function.
A local Lagrangian bias is introduced for initial, yet linear density fields as
\begin{equation}\label{defF}
1 + \delta_X(\vq) = F(\delta_R(\vq))
\end{equation}
where $\delta_X(\vq)$ is the overdensity of tracer $X$ and $\delta_R(\vq,t)$ is the  
initial density field linearly extrapolated up to time $t$ and smoothed by a window function 
over a scale $R_\Lambda$, $\delta_R(\vq) = \int d^3 q' W(|\vq - \vq'|/R_\Lambda)\delta_L(\vq',t)$. 
The bias can be made nonlocal in several ways, for example by promoting the function $F$ to  
a nonlocal functional \cite{Matsubara:2011ck,Matsubara:2012nc} or by including other operators as arguments \cite{Schmidt:2012ys,Vlah:2016bcl}. 
In Sec. \ref{Sect:3} we will add curvature and higher-derivative ($\nabla^{2} \delta_R, \nabla^{4} \delta_R, \cdots$)
arguments to $F$ with the purpose of removing $R_\Lambda$ dependencies on tracer statistics.
The choice of local Lagrangian bias leads inevitably to nonlocal Eulerian bias since
nonlinear evolution of smoothed fields is nonlocal. By the same reason an Eulerian local bias evolves into nonlocal bias; 
thus Eulerian local bias is not expected to hold in nature. 
Using tracer conservation, $(1+\delta_{X}(\vx))d^3x = (1+\delta_{X}(\vq))d^3q$, it is found that \cite{Matsubara:2008wx}
 \begin{align}\label{deltaX}
  (2\pi)^3\delta_\text{D}(\vk) + \delta_X(\vk) 
   &= \int d^3 q e^{-i\vk\cdot(\vq+\Psi(\vq))} \int  \frac{d\lambda}{2\pi}\tilde{F}(\lambda) e^{i \lambda \delta_R(\vq)},
 \end{align}
where $\tilde{F}(\lambda)$ is the Fourier transform of $F(\delta_R)$. We will call $\lambda$ the local bias spectral parameter. 
% If $F(\delta_R)=1+\delta_R \simeq 1 $, $F(\lambda)$ 
% becomes a Dirac delta function and we obtain the standard relation between the Lagrangian displacement and the matter overdensity. 
The power spectrum is
\begin{equation}\label{PX}
 (2\pi)^3 \delta_\text{D}(\vk) + P_X(k) = 
 \int d^3 q e^{i \vk \cdot \vq }\int \frac{d\lambda_1}{2\pi} \frac{d\lambda_2}{2\pi} \tilde{F}(\lambda_1) \tilde{F}(\lambda_2) 
 \langle e^{i[\lambda_1 \delta_1+\lambda_1 \delta_2 +  \vk\cdot \Delta]}\rangle ,
\end{equation}
where $\vq = \vq_2 - \vq_1$, $\delta_{1,2} = \delta_{R}(\vq_{1,2})$ and $\Delta_i = \Psi_i(\vq_2,t)-\Psi_i(\vq_1,t)$
is the difference of displacements at two Lagrangian coordinates.
With the aid of the cumulant expansion theorem we may write 
$\langle e^{ i X } \rangle = \exp \left( - \frac{1}{2} \langle X^2 \rangle_c - \frac{i}{6} \langle X^3 \rangle_c \right)$ which is valid up to third
order in fluctuations; in our case 
\begin{equation}\label{defX}
X =  \lambda_1 \delta_1+\lambda_2 \delta_2 + \vk\cdot \Delta.  
\end{equation}
We will further adopt the definitions \cite{Carlson:2012bu}
\begin{align} \label{defqfuncts}
 U^{mn}_i(\vq) = \langle \delta_1^m \delta_2^n \Delta_i \rangle_c, 
 \qquad A_{ij}^{mn}(\vq) =\langle \delta_1^m \delta_2^n \Delta_i \Delta_j\rangle_c, 
 \qquad W_{ijk}(\vq) =\langle \Delta_i \Delta_j \Delta_k\rangle_c, 
\end{align}
$U\equiv U^{01}=U^{10}$, $A_{ij}\equiv A^{00}_{ij}$,
and write
\begin{align}
-\frac{1}{2}\langle X^2 \rangle_c &=  
 -\frac{1}{2}k_ik_j A_{ij}  -\frac{1}{2}(\lambda_1^2 + \lambda_2^2) \sigma^2_R    
   - \lambda_1 \lambda_2 \xi_R - (\lambda_1 + \lambda_2) k_i U_i, \\
- \frac{i}{6} \langle X^3 \rangle_c &= -\frac{i}{2} (\lambda_1^2 +  \lambda_2^2) k_i U_i^{20}  
              -i \lambda_1\lambda_2 k_i U_i^{11} - \frac{i}{2} (\lambda_1 + \lambda_2) k_ik_j A_{ij}^{10} - \frac{i}{6} k_ik_jk_k W_{ijk},            
\end{align}
where we used $A_{ij}^{10} = A_{ij}^{01}$ and $U_i^{02} =U_i^{20}$. Explicit expressions for these \emph{$q$-functions} can be found in \cite{Carlson:2012bu}. 
% Fields $\delta_L(\vq,t)$ should be understood as initial Lagrangian matter density fields $\delta_R(\vq,t_{in})$ 
% linearly extrapolated up to the desired time, 
$\xi_R(q)$ is the correlation function of smoothed density fields and  $\sigma^2_R = \xi_R(0)$ their variance.
Lagrangian displacements, on the other hand, are not smoothed since they enter directly through the coordinate transformation of Eq.~(\ref{coordtrans})
% (in \cite{Kopp:2016crg} the authors find that one can follow better the center of mass displacement of proto-halos by also smoothing 
% the Lagrangian fields).
The strategy is to expand some terms out of the exponential,  
if we keep exponentiated the variances of matter smoothed overdensities we can introduce the bias parameters as \cite{Matsubara:2008wx} 
\begin{equation}\label{LagBiasDef}
  b_n \equiv \int_{-\infty}^{\infty} \frac{d\lambda}{2 \pi} \tilde{F}(\lambda)e^{-\lambda^2 \sigma^2_R/2}(i\lambda)^n 
\end{equation}
which will let us replace the $\lambda$ integrals for biases in 
Eq.~(\ref{PX}).\footnote{This approach is the same taken by Matsubara \cite{Matsubara:2008wx},  
but we write it here slightly different to generalize it in Sec. \ref{Sect:3}. Following the identity 
\begin{equation}
  \int_{-\infty}^{\infty} \frac{d\lambda}{2 \pi} \tilde{F}(\lambda)e^{-\lambda^2 \sigma^2_R/2}(i\lambda)^n =
  \frac{1}{\sqrt{2 \pi \sigma^2_R}}\int_{-\infty}^{\infty} d\delta e^{-\delta^2/2 \sigma^2_R} \frac{d^n F(\delta)}{d \delta^n} =\langle F^{(n)} \rangle
\end{equation}
we can identify $b_n=\langle F^{(n)} \rangle $ for Gaussian fields. 
} 
Forcing this definition to operate in Eq.~(\ref{deltaX}) we obtain
\begin{equation} \label{deltaB}
 (2\pi)^3 \delta_\text{D}(\vk) + \delta(\vk) = \int d^3q e^{-i \vk\cdot(\vq + \Psi)}
 \left(b_0+ (b_1-\frac{1}{2}\sigma_R^2 b_3)\delta_R(\vq) + \frac{1}{2}b_2 \big((\delta_R(\vq))^2 - \sigma^2_R \big)  
 +\frac{1}{6}b_3 (\delta_R(\vq))^3  +\cdots \right).
\end{equation}
On the other hand, the ``bare'' local bias parameters are given by \cite{Matsubara:2011ck}
\begin{align}\label{LagBarBiasDef}
 c_n\equiv\langle F^{(n)}(0) \rangle = \int_{-\infty}^{\infty} \frac{d\Lambda}{2 \pi} \tilde{F}(\lambda) (i \lambda)^n.
\end{align}
We can find a relation between the bare and renormalized biases by expanding the exponential in Eq.~(\ref{LagBiasDef}) and using
Eq.~(\ref{LagBarBiasDef})
\begin{equation} \label{cnTobn}
 b_n = \sum_{k=0}^{\infty} \frac{\sigma_R^{2k}}{2^k k!} c_{n+2k},
\end{equation}
from which we obtain standard relations $b_0=c_0 + \frac{1}{2} \sigma^2_R c_2 + \cdots$, 
$b_1=c_1 + \frac{1}{2}\sigma^2_R c_3 + \cdots$, $b_2 = c_2 + \cdots$, $b_3 = c_3 + \cdots$, where we neglected bias beyond third order.
Moreover, for Gaussian fields $b_0 = \langle F \rangle = 1$ and  we get a constraint equation
for even bare bias parameters, $\sum_{k=0}^{\infty} \frac{\sigma^{2k}}{2^k k!} c_{2k}=1$.
Hence, we interpret the renormalized bias expansion as a resummation of the unrenormalized biases that removes zero-lag correlators. 
Indeed, for initial density fields, such that $\Psi(q,t_{ini})=0$, from Eq.~(\ref{PX}) we obtain the
correlation function
% 
% Evaluating Eq.~(\ref{deltaB}) at initial time, such that $\Psi(\vq,t_{ini})=0$, we have
% \begin{equation} \label{btdf}
%  \delta_X(\vq,t_{ini}) =  b_1\delta_R(\vq,t_{ini}) + \frac{1}{2}b_2 \big((\delta_R(\vq,t_{ini}))^2 
%  - \sigma^2_R \big) +\frac{1}{6}b_3 (\delta_R(\vq,t_{ini}))^3 + \cdots
% \end{equation}
% recovering the most common expression for local Lagrangian  bias. 
% If we compute 
% the correlation function directly from Eq.~(\ref{btdf}) we obtain
% \begin{equation}
% \xi_X(\vq) = b_1(1+b_3\sigma^2_R)\xi_R + \frac{1}{2}b^2_2 \xi_R^2(\vq) + \cdots 
% \end{equation}
% and the bias parameters would require renormalization to remove dependences on the 
% smoothing scale that show up as zero-lag correlatores of the smoothed fields \cite{McDonald:2006mx}, 
% and in such a case one would prefer to use the letter `$c$' instead of `$b$' to denote them, following common usage. 
% An advantage of introducing the bias through  Eq.~(\ref{LagBiasDef}),
% is that zero-lag correlators are automatically out of tracers statistics. Indeed, again for initial density fields, by setting 
% $\Psi(\vq,t_{ini})=0$ in Eq.~(\ref{PX}), we obtain the correlation function with local renormalized bias 
\begin{equation} \label{xiXL2}
 \xi_{X,L}(\vq) = \sum_{n=1}^\infty \frac{b_n^2}{n!} (\xi_{R,L}(\vq))^n, 
\end{equation}
which has no zero-lag correlators.
The label ``$L$'' means that at the end of the process we have evolved linearly the correlations in the right-hand side (rhs) of the above
equation, and not that $\xi_{X,L}(\vq)$ is the linear correlation function for tracers. That is, the theory is regulated by
two scales, the scales of nonlinearity $k_{NL}$ for fluctuations, and the scale $\Lambda=1/R_{\Lambda}$ associated to the bias expansion. 
Hereafter, we will suppress that label under the understanding that terms composed of smoothed fields evolve linearly, and we
use it only to distinguish between the linear and loop contributions of quantities. 
Now, allowing nonlinear evolution of Lagrangian displacements in Eq.~(\ref{PX}) and using Eqs.~(\ref{defqfuncts}) we have
\begin{align}\label{XLPTPS}
 (2\pi)^3 \delta_\text{D}(\vk) + P_X^\text{LPT}(k) &= \int d^3 q e^{i \vk \cdot \vq } e^{ -\frac{1}{2}k_ik_j A_{ij} - \frac{i}{6} k_ik_jk_k W_{ijk} }
   \Bigg[ 1  + b_1^2 \xi_R + 2 i b_1 k_i U_i + \frac{1}{2} b_2^2 \xi_R^2   \nonumber\\
  -(b_2  &+ b_1^2) k_i k_j U_i U_j
  + 2i b_1 b_2 \xi_R k_i U_i 
  + i b_1^2 k_i U_i^{11} + i b_2 k_i U_i^{20}- b_1 k_ik_j A_{ij}^{10} \Bigg].
\end{align}
This is the exact expression for the 1-loop LPT power spectrum with a second order local bias expansion.
Since the exponential is highly oscillatory it is challenging to numerically solve the integral, this has been done
for matter  in \cite{Sugiyama:2013mpa,Vlah:2014nta} adopting different methods.  
The idea of CLPT is to perform a further expansion keeping only
quadratic terms in $k$ in the exponential, in such a way that one can perform the Fourier transform and get an analytical expression for 
the correlation function by performing several
multivariate Gaussian integrals. Different schemes are possible, but in order to preserve Galilean invariance these reduce to essentially two. 
In Ref.~\cite{Carlson:2012bu}, the contribution $A=A_L + A_{loop}$ is kept exponentiated while $W$ is expanded. 
In order to treat in equal footing linear and
loop contributions we follow \cite{Vlah:2015sea} and expand also the nonlinear piece $A_{loop}$. By doing this to Eq.~(\ref{XLPTPS}) and
Fourier transforming we obtain
\begin{align}\label{XCLPTCF}
1 + \xi_X^\text{CLPT}(r) &= \int  \frac{d^3 q}{(2 \pi)^{3/2} |A_L|^{1/2}} e^{- \frac{1}{2}(A_L^{-1})_{ij}(q_i-r_i)(q_j-r_j) } 
\Bigg( 1 - \frac{1}{2} A_{ij}^{loop}G_{ij} +\frac{1}{6}\Gamma_{ijk}W_{ijk}  \nonumber\\
   & \qquad - b_1 (2 U_i g_i + A^{10}_{ij}G_{ij}) + b_1^2 (\xi_R - U_iU_jG_{ij}- U_i^{11}g_i)  - b_2 (U_i^{20}g_i + U_iU_jG_{ij})    \nonumber\\
   & \qquad  - 2 b_1 b_2 \xi_R U_i g_i + \frac{1}{2}  b_2^2  \xi_R^2  \Bigg), 
\end{align}
with
\begin{equation}
g_{i}= (A_L^{-1})_{ij} (q_j - r_j),\qquad G_{ij} = (A_L^{-1})_{ij} - g_i g_j,\qquad \Gamma_{ijk} = (A_L^{-1})_{\{ij}g_{k\}} + g_i g_j g_k, 
\end{equation}
taking the form of a Gaussian convolution. 
Indeed, the integrand at large fixed $r$ is very close to a Gaussian centered at $q=r$ with a width $\sim 20 \,\text{Mpc}/h$. 
% thus the correlation function
% has the appealing form of the convolution of two Gaussians \emph{(checar!)}. 
A nice feature of the CLPT correlation function is that it preserves the Zel'dovich approximation as its lower order contribution, corresponding to
the ``1'' in between the parentheses. 
The different contributions to the above equation are plotted in Fig.~\ref{fig:BiasComp} at redshift $z=0$.

In Ref.~\cite{Carlson:2012bu}, by using Eq.~(\ref{XCLPTCF}) directly, it was shown that the linear correlation function for tracers is
\begin{equation} \label{xiXL}
 \xi_X(r) = (1+b_1)^2 \xi_L(r),
\end{equation}
following the identification $b_1^E = 1+b_1$, where $E$ refers to the Eulerian bias.
This result is in apparent contradiction with Eq.(\ref{xiXL2}). Nevertheless, by allowing linear evolution of the 
Lagrangian displacement in Eq.~(\ref{deltaB}), we find\footnote{This can be done by noting 
\begin{equation}
 \int d^3q e^{i\vk\cdot(\vq + \Psi)} = \int J^{-1} d^3 x e^{i\vk\cdot \vx} = \int d^3 x e^{i\vk\cdot \vx} (1-\Psi_{i,i} + \cdots),
\end{equation}
where $J$ is the Jacobian determinant of the coordinate transformation, Eq.~(\ref{coordtrans}).
}
$\xi_{X}(r) =  b_1^2 \langle \delta_R(\vx+\ve r)\delta_R(\vx) \rangle 
+ 2 b_1  \langle \delta_R(\vx+\ve r)\delta(\vx) \rangle +  \langle \delta(\vx+\ve r)\delta(\vx) \rangle$, recovering Eq.~(\ref{xiXL})
at scales we can neglect the smoothing, ideally this is for $r > R_\Lambda$, but we will find this inequality to be more restrictive. 
Some works attach the smoothing filter to the bias by defining $b_1(k) = b_1 \tilde{W}(k R)$ 
in Fourier space \cite{Matsubara:2011ck,Matsubara:2012nc}; 
that approach makes Eq.~(\ref{xiXL}) consistent at any scale beyond $R_\Lambda$.

% This is 
% modified by the bias terms as $\xi_X^\text{ZA}(r) = (1+\langle F'\rangle)^2 \xi^\text{ZA}(r)$ \emph{(Fake. checar!)}
\begin{figure}
	\begin{center}
	\includegraphics[width=3 in]{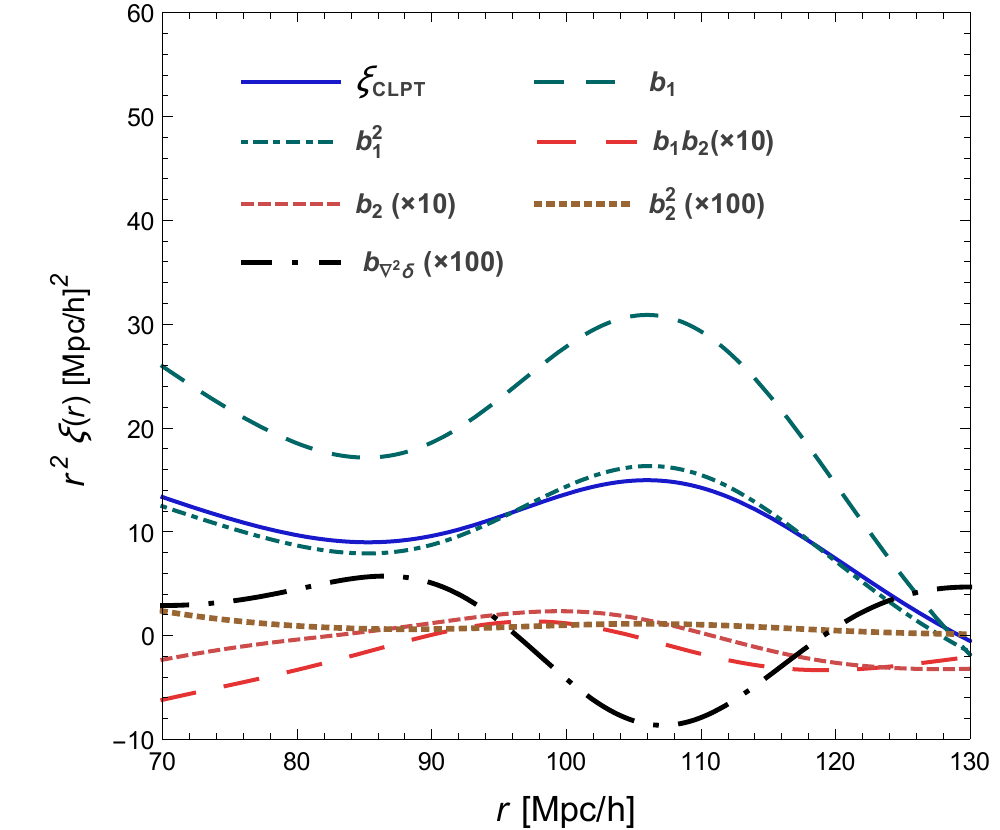}
	\includegraphics[width=3 in]{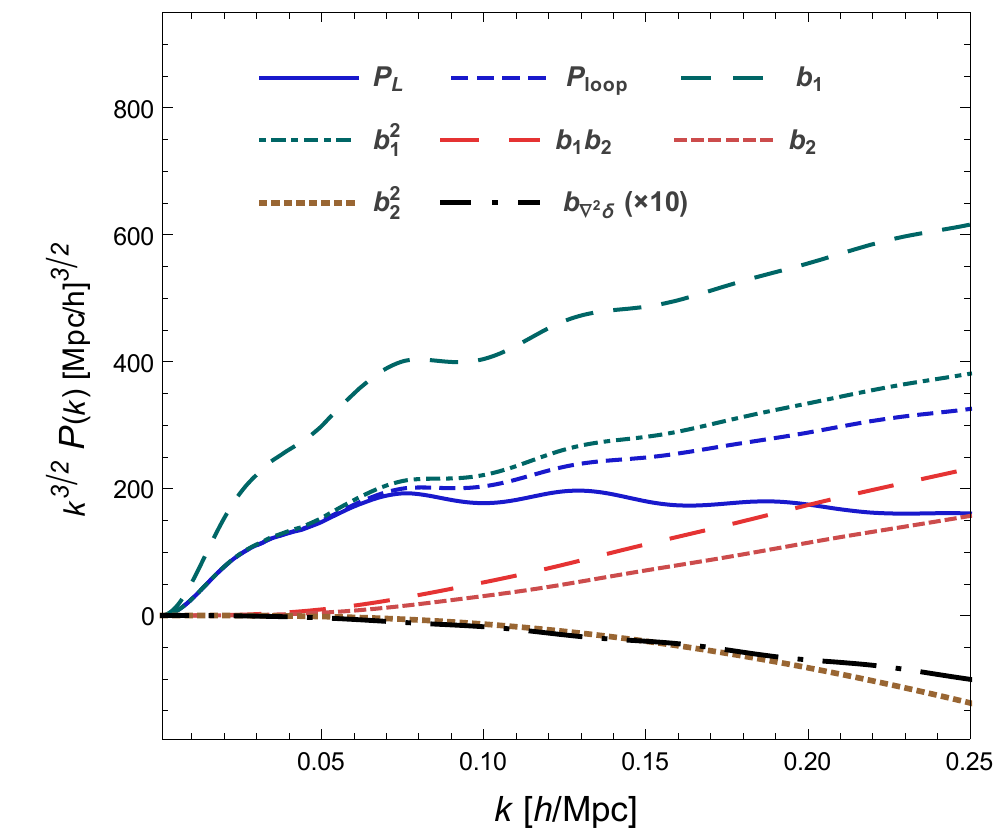}	
	\caption{Bias components for the CLPT correlation function [Eq.~(\ref{XCLPTCF})] and SPT power spectrum [Eq.~(\ref{XSPTPS})]. We are also showing
	the contribution of curvature bias with bias parameter $b_{01}=b_{\nabla^2\delta}$ of Sects. \ref{Sect:3} and \ref{Sect:4}. For the correlation function
	this is the term that contains $c_{10}c_{01}\nabla^2\xi_R$ in Eq.~(\ref{CLPTCFgradD}), the second term in that equation is degenerated with this one,
	while the rest are subdominant. For the power spectrum we are showing $-k^2 P_L(k)$.
	 We fix cosmological parameters to the best fit of the WMAP nine-year results \cite{2013ApJS..208...19H} and consider redshift $z=0$. 
	\label{fig:BiasComp}}
	\end{center}
\end{figure}

The SPT power spectrum is obtained by expanding all terms out of the exponential in Eq.~(\ref{XLPTPS}) and by performing the $q$ integral,
obtaining \cite{Matsubara:2008wx}
\begin{align}\label{XSPTPS}
  P^\text{SPT}_X(k)  &=  P_L(k)  + P_{22}(k) + P_{13}(k) + b_1 a_{10}(k) + b_2 a_{01}(k)  + b_1^2 a_{20}(k)
   + b_1 b_2 a_{11}(k) + b_2^2 a_{02}(k), 
\end{align}
with
\begin{align}
  P_{22}(k) &=  \frac{9}{98}Q_1(k)+ \frac{3}{7}Q_2(k) + \frac{1}{2}Q_3(k),\qquad   
  P_{13}(k) =  \frac{10}{21}R_1(k) + \frac{6}{7}R_2(k) - \sigma^2_L k^2 P_L(k),\label{P22P13}\\
  a_{10}(k) &= 2 P_L(k) + \frac{10}{21}R_1(k) +\frac{6}{7} R_{1+2}(k) +\frac{6}{7} R_{2}(k) + \frac{6}{7} Q_{5}(k) + 2(Q_7(k)-\sigma^2_L k^2 P_L(k)),   \\
  a_{01}(k) &=  Q_{9}(k) + \frac{3}{7}Q_8(k),\qquad a_{20}(k) = P_L(k)  + \frac{6}{7}R_{1+2}(k)  + Q_9(k)+ Q_{11}(k)- \sigma^2_L k^2 P_L(k),\\
  a_{11}(k) &= 2 Q_{12}(k),\qquad a_{02}(k) = \frac{1}{2} Q_{13}(k), \\
  \sigma^2_L &= \frac{1}{3}\delta_{ij} \langle \Psi_i(0)\Psi_j(0) \rangle = \frac{1}{6\pi^2}\int dp P_L(p).
\end{align} 
In Fig.~\ref{fig:BiasComp} we show the different contributions to Eq.~(\ref{XSPTPS}).
% and $\sigma^2_L$ is one third the trace of the variance of linear Lagrangian displacements: 
% and   %Equation (\ref{XSPTPS}) is derived in appendix ?.
% So far, these results are general and were derived in \cite{Matsubara:2007wj,Matsubara:2008wx,Carlson:2012bu}. 
% In generalized cosmologies, though, the functions $Q(k)$ and $R(k)$ are different than in $\Lambda$CDM.
The $Q(k)$ and $R(k)$ functions are computed for Einstein-de Sitter (EdS) evolution in \cite{Matsubara:2007wj,Matsubara:2008wx}, however, they
can differ in more general cosmologies \cite{Aviles:2017aor}. In particular, in $\Lambda$CDM we have
that $R_{1}+R_{2} \simeq R_{1+2} $ is a good approximation, holding exactly in EdS.
The claim that Eq.~(\ref{XSPTPS}) contains the standard pieces in the unbiased SPT power spectrum was proven in  \cite{Matsubara:2007wj} 
for EdS and more generally in \cite{Vlah:2014nta}.

We notice Eq.~(\ref{XSPTPS}) is the SPT power spectrum for locally biased tracers at initial time, 
which we emphasize differs from the power spectrum of Eulerian local biased tracers, because smoothing and nonlinear evolution do not commute.
Only at linear order in fluctuations these are simple related since 
$P_{XL}(k) = (1+ b_1)^2 P_L(k)$, consistent with Eq.~(\ref{xiXL}).
However, at large scales the biased power spectrum does not reduce to the linear one, instead we have
\begin{equation}
P_{X}^\text{SPT}(k\rightarrow 0) = (1+b_1)^2 P_L(k) + \frac{1}{2} b_2^2 \int \Dk{p} (P_L(p))^2.
\end{equation}
% We can trace back the appeareance of the constant term to the splitting of $\langle X \rangle_c$ above. Since
% $\langle X^2 \rangle_c(q=0) = -\frac{1}{2}(\lambda_1 + \lambda_2)^2\sigma^2_R$, we may be tempted to retain it in the exponential  of Eq.~(), before
% the $\lambda$ integration. But by doing so, we find zero-lag correlators in the expansion, ruining the  
% advantage of the bias definition [Eq.~()].
% This procedure, though, is infructuous since it leads to negative 
% contributions for bias $\langle F'\rangle$ and  $\langle F'\rangle^2$ 
% in the correlation function.
% Discussion.... See  [R.J.Scherrer and D.H.Weinberg, Astrophys.J., 504, 607(1998)], see also Halo Zeldovich... Matsubara2008b
% In astro-ph/0609413 Eq.10 this term exactly appear.
The constant term arises from the last term in Eq.~(\ref{XSPTPS}), 
\begin{equation} \label{a02div}
a_{02}(k) = \frac{1}{2}\int \Dk{p} P_L(p) P_L(|\vk-\vp|),
\end{equation}
which is potentially harmful.
For power law power spectra with $P_L \propto k^n$, $a_{02}(k)$ is UV divergent for $n\geq-3/2$ 
(instead of $P_{loop}$, whose UV divergence appears for $n\geq -1 $). The filtering
makes $a_{02}(k)$ convergent but sensible to the cutoff $\Lambda=1/R_\Lambda$.
Thus, scales below $\Lambda$ receive arbitrary corrections from the small scales that were integrated out of the 
theory.
% $R_\Lambda$-dependence of results is shared by Eulerian treatment of local bias, leading to the renormalization
% of bias parameters \cite{McDonald:2006mx}. Indeed, 
Exactly the same constant contribution to the power spectrum is present in the biased SPT power spectrum 
with already renormalized Eulerian bias, that is cured by considering 
the addition of a constant shot noise to the biased power spectrum  \cite{McDonald:2006mx}, 
which at this point we introduce as $P_X \rightarrow P_X + N_0$. This white noise arises from 
stochastic, uncorrelated with long wavelength overdensities, contributions to the density fields of tracers, and is renormalized to absorb
the term $a_{02}(k=0)$. Slightly more formally, in Eq.~(\ref{defX}) we may add a contribution $X_\epsilon=\lambda_{\epsilon,1} \epsilon_0(\vq_1)+\lambda_{\epsilon,2} \epsilon_0(\vq_2)$ that comes from
stochasticity of small scales, inducing the constant $N_0 = b_{\epsilon}^2 \langle \epsilon_0(\vk) \epsilon_0(\vk') \rangle$ 
contribution to the power spectrum (see Sec. \ref{Sect:4}). Thereafter, the bias parameter $b_{\epsilon}$ absorbs the constant $a_{02}(k=0)$, leading
to a renormalized function
\begin{equation}\label{rena02}
 a_{02}(k) \longrightarrow  a_{02}(k) - a_{02}(0) = \frac{1}{2} \int \Dk{p} P_L(p) (P_L(|\vk-\vp|) - P_L(p)), 
\end{equation}
which is safe from UV divergences for $n < -1/2$. By taking the inverse Fourier transform of the renormalized 
function $a_{02}(k)$, the constant shift only contributes with a Dirac delta function at $r=0$
in the correlation function.

There are other divergences in Eq.~(\ref{XSPTPS}). $Q_{3}(k)$ has IR divergences when the internal momentum goes to zero and
when it approaches the external momentum; both divergences are canceled by the term $2\sigma^2_L k^2 P_L$  \cite{Aviles:2017aor}. Functions 
$Q_7(k)$ and $Q_{11}(k)$ present the same IR divergence when the internal momenta go to zero, 
but we note that they are accompanied by $\sigma^2_L k^2 P_L$ terms; hence these divergences cancel out, 
becoming IR safe for spectral index $n>-3$. 
All the other functions in Eq.~(\ref{XSPTPS}) are well behaved.  

\end{section}

\begin{section}{Curvature bias and renormalization}\label{Sect:3}

\begin{subsection}{Density curvature bias}\label{Sect:3_1}

To some extent, we have removed the $R_\Lambda$ dependence from statistics in the sense that they lack zero-lag correlators. 
However, there still exist some 
residual dependencies. Consider the linear correlation function of smoothed density fields \cite{Schmidt:2012ys}
\begin{equation} \label{xiFromxiR}
 \xi_R(q) = \int \Dk{k} e^{i\vk\cdot\vq} |\tilde{W}(kR_\Lambda)|^2 P_L(k) =  \xi(q) + 2 R_\Lambda^2\nabla^2 \xi(q) + \mathcal{O}(R_\Lambda^4\nabla^4 \xi(q)) ,
\end{equation}
where we expanded the smoothing filter as $\tilde{W}(kR_\Lambda) = 1 - R_\Lambda^2k^2 +  \cdots$ for illustration purposes, 
though our results do not depend
in this particular choice. Equation (\ref{xiFromxiR}) shows that the linear correlation function will be $R_\Lambda$ independent as long as
$R^2_\Lambda\nabla^2 \xi(r) \ll  \xi(r)$. For a featureless correlation function this holds as long as $R_\Lambda\ll r$, 
but in our Universe where the BAO bump with a width $\Delta r\sim 20\, \text{Mpc}/h$ is present at a scale $r\sim 100 \, \text{Mpc}/h$, the
condition becomes $ R_\Lambda \ll \Delta r$, which is highly undesirable, especially for the description of massive halos where
the smoothing scale is typically identified with  its Lagrangian radius. 
In this section, following closely the work of Ref.~\cite{Schmidt:2012ys}, we 
remove the subleading scale dependencies by introducing a density Laplacian as an argument in the Lagrangian bias function,
\begin{equation}\label{defF2}
  1+\delta_X = F(\delta_R,\nabla^2 \delta_R),
\end{equation}
which generalizes the LPT power spectrum for tracers, 
\begin{equation}\label{PX2}
 (2\pi)^3 \delta_\text{D}(\vk) + P_X(k) = \int d^3 q e^{i \vk \cdot \vq }\int \frac{d^2\mathbf{\Lambda}_1}{(2\pi)^2} 
 \frac{d^2\mathbf{\Lambda}_2}{(2\pi)^2} \tilde{F}(\mathbf{\Lambda}_1) \tilde{F}(\mathbf{\Lambda}_2) 
 \langle e^{i[\mathbf{\Lambda}_1 \cdot \mathbf{D}_1 +\mathbf{\Lambda}_2 \cdot \mathbf{D}_2  +  \vk\cdot \Delta]}\rangle ,
\end{equation}
where 
\begin{equation}
\mathbf{\Lambda} = (\lambda,\bar{\eta}), \qquad \mathbf{D} = (\delta_R,\nabla^2\delta_R)
\end{equation}
are vectors and $\tilde{F}(\mathbf{\Lambda})= \tilde{F}(\lambda,\bar{\eta})$ is the Fourier transform of $F(\delta_R,\nabla^2\delta_R)$ with respect to both
arguments. In the same way that $\lambda$ is the 
spectral bias parameter of matter overdensities, $\bar{\eta}$ is the (bare) spectral bias 
parameter for the curvature operator $\nabla^2\delta_R$. 
We introduce the bivariate bias parameters as a generalization of Eq.~(\ref{LagBiasDef}):
\begin{equation}\label{LagBiasDef2}
 c_{nm} \equiv \int \frac{d^2\mathbf{\Lambda}}{(2 \pi)^2} \tilde{F}(\mathbf{\Lambda})
 e^{-\frac{1}{2}\mathbf{\Lambda}^\text{T} \mathbf{\Sigma} \mathbf{\Lambda}}(i\lambda)^n (i \bar{\eta})^m =
 \int \frac{d^2 \mathbf{D}}{2 \pi |\mathbf{\Sigma}|^{1/2}} e^{-\frac{1}{2}\mathbf{D}^\text{T} \mathbf{\Sigma}^{-1} \mathbf{D}} 
 \frac{\partial^{n+m} F(\delta_R,\nabla^2 \delta_R)}{\partial \delta^n \partial (\nabla^2\delta)^m} 
 = \left\langle \frac{\partial^{n+m} F(\delta_R,\nabla^2 \delta_R)}{\partial \delta^n \partial (\nabla^2\delta)^m} \right\rangle. 
\end{equation}
The components of the covariance matrix are given by 
zero-lag correlators as $\Sigma_{11} = \langle \delta_R^2 \rangle = \sigma^2_R$,
$\Sigma_{12} = \Sigma_{21} = \langle \delta_R \nabla^2 \delta_R \rangle$, and $\Sigma_{22} = \langle (\nabla^2 \delta_R)^2 \rangle$.
Standard notation is recovered by identifying 
\begin{equation}
b_n=c_{n0}, \qquad \text{and} \qquad c_{(\nabla^{2}\delta)^m} = c_{0m}.
\end{equation}
We see below that parameters $c_{nm}$ require renormalization, unlike the $b_n$ of the previous section. For initial density fields, we get 
the correlation function for tracers
\begin{equation} \label{xiXBibias}
 \xi_X(\vq) = c_{10}^2 \xi_R(\vq) + 2 c_{10}c_{01} \nabla^{2} \xi_R(\vq)  + c_{01}^2 \nabla^{4}\xi_R(\vq),
\end{equation}
which extends Eq.~(\ref{xiXL2}) by including curvature bias, but simplifies it by considering only linear fluctuations. 
It is good to keep in mind 
that $c_{nm}$ has units of $[\text{\emph{length}}]^{2m}$, reflecting the nonlocality of the bias description.

In this subsection, we are interested in removing the $R_\Lambda$ dependencies of the $\xi_R$ and $\xi_R^2$ terms 
of Eq.~(\ref{XCLPTCF}).\footnote{Contrary to $\xi_R$, $U_i$ and $A^{10}_{ij}$
functions are sufficiently smooth at large scales, such that when expanded analogously to Eq.~(\ref{xiFromxiR}), 
terms as $R_\Lambda^2 \nabla^2 U_i$ can be neglected for $q> R_{\Lambda}$.}
This can be achieved by considering the following contributions to the LPT 
power spectrum\footnote{The whole second order bias expansion, including all second order terms
$c_{10}$, $c_{01}$,$c_{20}$, $c_{11}$ and $c_{02}$ is computed in Appendix \ref{app:BiasExp}. 
Equation (\ref{LPTPSgradD}) is a subset of Eq.~(\ref{Bias2orderExp}).}
 \begin{align} \label{LPTPSgradD}
 (2\pi)^3 \delta_\text{D}(\vk) + P_X^\text{LPT}(k) &\supset  \int d^3 q e^{i \vk \cdot \vq } e^{ -\frac{1}{2}k_ik_j A_{ij} -\frac{i}{6}k_ik_j k_k W_{ijk} }
   \Big(2 c_{10} c_{01} \nabla^2\xi_R(q) - 2 i c_{01} k_i \nabla_i \xi_R(q) \nonumber\\
   &\quad +2i (c_{20}c_{01} + c_{10}c_{11} )\nabla^2 \xi_R k_i U_i + 2 c_{20}c_{11}\xi_R \nabla^2 \xi_R \Big),
\end{align}
with
 \begin{align}
 \nabla^2\xi_R(q)  
 = -\int \Dk{k} e^{i \vk \cdot \vq} k^2 |\tilde{W}(kR)|^2 P_L(k) = \nabla^2 \xi(q) + 2R_\Lambda^2 \nabla^4 \xi(q) + \cdots,  %\\
%  -\nabla_i \xi_R(q)  
%   = \hat{q}_i \int \frac{dk}{2 \pi^2} k^3 \tilde{W}(kR) P_L(k) j_1(kq) = -\nabla_i \xi(q) - R_\Lambda^2 \nabla_i \nabla^2 \xi(q) + \cdots 
\end{align}
where we expanded the filter $W$ in powers of $R_\Lambda^2k^2$.
By expanding loop contributions out of the exponential in Eq.~(\ref{LPTPSgradD}) and performing the Fourier transform, we arrive at 
\begin{align}\label{CLPTCFgradD}
 \xi_X^\text{CLPT}(r) &\supset \int  \frac{d^3 q}{(2 \pi)^{3/2} |A_L|^{1/2}} e^{- \frac{1}{2}(A_L^{-1})_{ij}(q_i-r_i)(q_j-r_j) } 
  \Big(  2 c_{10} c_{01} \nabla^2 \xi_R(q) + 2  c_{01} \nabla_i \xi_R(q) g_i \nonumber\\
    &\quad -2(c_{20}c_{01}+c_{10}c_{11})\nabla^2 \xi_R g_i U_i + 2 c_{20}c_{11}\xi_R \nabla^2 \xi_R  + \cdots \Big).
\end{align}
A similar result is presented in \cite{Vlah:2016bcl}, where the authors additionally consider tidal bias and EFT contributions; the latter are degenerated with
the curvature bias. 
By comparing Eqs.~(\ref{xiXBibias}) and (\ref{CLPTCFgradD}) we note an interesting fact: once nonlinear evolution takes place, 
bivariate bias parameters $c_{nm}$, with both $n\neq 0$ and $m\neq 0$, should be 
considered, and the description with only univariate biases $c_{n0}$ and $c_{0m}$ becomes incomplete. 
Clearly, this feature is shared by the SPT power spectrum, but we postpone its
discussion to Sec.~\ref{Sect:4}.

Joining this result with the $\xi_R$ in Eq.~(\ref{XCLPTCF}) and using Eq.~(\ref{xiFromxiR}), 
the combination
\begin{equation}\label{renGrad1}
 c_{10}^2 (\xi(q) + 2 R_\Lambda^2\nabla^2 \xi(q)) + 2 c_{10} c_{01} \nabla^2 \xi(q) = c_{10}^2\xi(q) + 2 c_{10}(c_{10} R_\Lambda^2 + c_{01}  )\nabla^2 \xi(q)
\end{equation}
appears in the correlation function. That is,  the curvature bias operator has introduced the precise term 
in order to absorb the $R_\Lambda^2$ term in the above equation, since we can reparametrize $b_{01} =  c_{01}+c_{10} R_\Lambda^2$. 
In this sense, the addition of a curvature bias is not a choice, it should be included to make the theory 
independent of (or less sensible to) the details of the smoothing. 
% In the expansion of Eq.~(\ref{xiFromxiR}) we found terms that behave as $R^{2n} \nabla^{2n} \xi$, 
% their smoothing scale dependence can be removed with the introduction of higher derivative bias, as 
% shown in the upcoming section; see also appendix B of \cite{Schmidt:2012ys}.

Another contribution to Eq.~(\ref{XCLPTCF}) sensible to $R_\Lambda$ is $\xi_R g_i U_i$. We can pair it with the term $\nabla^{2}\xi_R g_i U_i$ 
in Eq.~(\ref{CLPTCFgradD}), leading to
\begin{equation}\label{renGrad2}
 \left[2 c_{10}c_{20} (\xi + 2 R_\Lambda^2\nabla^2 \xi)  + 2  (c_{20}c_{01}+ c_{10} c_{11}) \nabla^2 \xi \right]  g_i U_i
 = \left[ 2c_{10}c_{20} \xi  +   2  c_{20}(2 c_{10} R_\Lambda^2 + c_{01}+ \frac{c_{11}c_{10}}{ c_{20}} ) \nabla^2 \xi \right] g_i U_i. 
\end{equation}
One of the terms $c_{10} R^2$ is absorbed by $c_{01}$ as in Eq.(\ref{renGrad1}), while the other is absorbed by $\frac{c_{11}c_{10}}{ c_{20}}$.
That is, we can reparametrize $b_{11}= c_{11} + c_{20}R_\Lambda^2$.

Analogously, the contribution $\frac{1}{2}c_{20}^2 \xi^2_R$ to Eq.~(\ref{XCLPTCF}) is expanded as
\begin{equation}
\frac{1}{2}c_{20}^2 \xi^2_R = \frac{1}{2}c_{20}^2 \xi^2 + 2 c_{20}^2\xi R_\Lambda^2  \nabla^2 \xi  + \mathcal{O}(R_\Lambda^4 \nabla^4 \xi). 
\end{equation}
We join up the second term in the rhs of the above equation with the term $2 c_{20} c_{11} \xi_R \nabla^2 \xi_R$ in Eq.~(\ref{CLPTCFgradD}), the sum of both is
\begin{equation}\label{renGrad3}
 2 \frac{c^2_{20}}{c_{10}}\left(c_{10} R_\Lambda^2 + \frac{c_{11}c_{10}}{ c_{20}}\right)\xi  \nabla^2 \xi + \cdots,
\end{equation}
and consistently with Eq.~(\ref{renGrad2}), the term $\frac{c_{11}c_{10}}{ c_{20}}$ absorbs $c_{10} R_\Lambda^2$.

\end{subsection}

\begin{subsection}{Renormalization of curvature and higher order bias via spectral parameters}\label{Sect:3_2}

The renormalization presented in the previous subsection is a special case of a more general method
that is the subject of this subsection. 
We can guarantee that our results are $R_\Lambda$ independent if we are able to remove any 
$R_\Lambda$ dependence in the function
$X$ [Eq.~(\ref{defX})]. We focus on just one term, $X_\delta = \lambda \delta_R$, and use
\begin{equation} \label{expdR}
 \delta_R(\vq) = \int \Dk{k} e^{i\vk\cdot\vq} \tilde{W}(kR_\Lambda) \delta(\vk) =  \sum_{n=0}^\infty (-1)^n W_n R_\Lambda^{2n}\nabla^{2n}\delta(\vq) 
\end{equation}
where we assumed that the filter can be Taylor expanded as $\tilde{W}(kR) = \sum_{n=0}^\infty W_n (kR_\Lambda)^{2n}$; we notice that 
normalization of the window function, $\int d^3x W(\vx) = 1$, implies $W_0=1$. To absorb the smoothing scale
we add a (formally infinite) collection of counterterms
\begin{align} \label{Yfunction}
 Y= \sum_{m=1}^{\infty} \bar{\eta}_{\nabla^{2m}\delta} \nabla^{2m}\delta_R(\vq).
\end{align}
where $\bar{\eta}_{\nabla^{2m}\delta}$ is a set of unrenormalized spectral parameters; 
in the notation of the previous subsection $\bar{\eta}= \bar{\eta}_{\nabla^{2}\delta}$. 
% If we cut the sum at $m=N$, we keep 
% dependences $ \sim \mathcal{O}(R^{2(N+1)}\nabla^{2(N+1)}\delta_R)$. 
Inserting Eq.~(\ref{expdR}) in Eq.~(\ref{Yfunction}) and summing to $X_\delta$ we get
% \begin{align}
% X_\delta + Y &= \lambda  \delta(\vq) + \sum_{n=1}^{\infty} \left( (-1)^n \lambda W_n R_\Lambda^{2n} +
% \sum_{m=1}^{n}
% (-1)^{n+m} W_{n-m} R_\Lambda^{2(n-m)}  \bar{\eta}_{\nabla^{2m}\delta} \right) \nabla^{2n} \delta(\vq) \nonumber\\
% &= \lambda \delta(\vq) + \sum_{n=1}^{\infty} \eta_{\nabla^{2n}\delta}\nabla^{2n}\delta(\vq),
% \end{align}
% where we introduced the renormalized bias spectral parameters
% \begin{equation} \label{spectralbiasrel}
%  \eta_{\nabla^{2n}\delta} = (-1)^n W_n R_\Lambda^{2n} 
%  \left( \lambda + \sum_{m=1}^n (-1)^m \frac{W_{n-m}}{W_n R_\Lambda^{2m}}\bar{\eta}_{\nabla^{2m}\delta}  \right).
% \end{equation}
\begin{align}
X_\delta + Y &= \lambda  \delta(\vq) + \sum_{n=1}^{\infty} \left( (-1)^n \lambda W_n R_\Lambda^{2n} +
\sum_{m=1}^{n}
(-1)^{n-m} W_{n-m} R_\Lambda^{2(n-m)}  \bar{\eta}_{\nabla^{2m}\delta} \right) \nabla^{2n} \delta(\vq) 
\end{align}
where we used the double sum identity $\sum_{m=1}^{\infty}\sum_{n=0}^{\infty} = \sum_{i=1}^{\infty}\sum_{m=1}^{i}$ with $i=m+n$, and relabel $i\rightarrow n$.

We introduce the renormalized bias spectral parameters as
\begin{equation} \label{spectralbiasrel}
 \eta_{\nabla^{2n}\delta} = (-1)^n W_n R_\Lambda^{2n} 
 \left( \lambda + \sum_{m=1}^n (-1)^m \frac{W_{n-m}}{W_n R_\Lambda^{2m}}\bar{\eta}_{\nabla^{2m}\delta}  \right),
\end{equation}
from which
\begin{align}
X_\delta + Y
&= \lambda \delta(\vq) + \sum_{n=1}^{\infty} \eta_{\nabla^{2n}\delta}\nabla^{2n}\delta(\vq),
\end{align}
becomes $R_\Lambda$ independent. 

We have introduced an infinite set of counterterms; in this case zero-lag correlators 
in the covariance matrix $\mathbf{\Sigma}$ diverge for common linear power spectra. 
In practice, only a finite set of counterterms can be introduced, let us say up to 
$n=N$, keeping  dependencies $ \sim \mathcal{O}(R_\Lambda^{2(N+1)}\nabla^{2(N+1)}\delta_R)$, and as one approaches the smoothing scale, the theory
loses its validity.

We now come back to the case in which only the counterterm $\nabla^2\delta_R$ is introduced. 
From Eq.~(\ref{spectralbiasrel}), the spectral renormalized parameter $\eta=\eta_{\nabla^2 \delta}$ is
\begin{equation} \label{spectralbiasrelLap}
\eta =  \bar{\eta} - W_1 R_\Lambda^2 \lambda. 
\end{equation}
We define the bivariate renormalized bias parameters as
\begin{equation}\label{DefBivRenPar}
 b_{nm} \equiv \int \frac{d \mathbf{\Lambda}}{(2 \pi)^2} e^{-\frac{1}{2} \mathbf{\Lambda}^T \mathbf{\Sigma} \mathbf{\Lambda}}
 \tilde{F}(\mathbf{\Lambda}) (i \lambda)^n (i \eta)^m,
\end{equation}
where we note that we still have $\mathbf{\Lambda} = (\lambda, \bar{\eta})$. Replacing $\bar{\eta}$ for $\eta$, we get
\begin{equation}\label{bivRenToBar}
b_{nm} = \sum_{k=0}^m \binom{m}{k} (-W_1)^{k}R_\Lambda^{2k} c_{n+k,m-k},
\end{equation}
which relates renormalized and unrenormalized bivariate biases. We immediately have $b_{n0}=c_{n0}$ and
\begin{align}
b_{01} = c_{01} - W_1 R_\Lambda^2 c_{10},\qquad b_{11} = c_{11} - W_1 R_\Lambda^2 c_{20}. 
\end{align}
By setting $W_1=-1$, as in the previous subsection, we note that these are the 
precise relations we need to make the $R_\Lambda$ cancellations in Eqs.~(\ref{renGrad1}), (\ref{renGrad2}) and (\ref{renGrad3}). 

We can extend the definitions of bivariate bias in Eqs.(\ref{LagBiasDef2}) and (\ref{DefBivRenPar}) to include higher-derivative terms.
That is, we introduce the bare and renormalized multivariate bias parameters as
\begin{equation} \label{barbiasparamDef}
 c_{n_0n_1\cdots n_N} = \int \frac{d\mathbf{\Lambda}}{(2\pi)^{N+1}} e^{-\frac{1}{2}\mathbf{\Lambda}^\text{T}\mathbf{\Sigma}\mathbf{\Lambda}}
 \tilde{F}(\mathbf{\Lambda})(i\lambda)^{n_0}(i\bar{\eta}_{\nabla^2\delta})^{n_1} \cdots (i\bar{\eta}_{\nabla^{2N}\delta})^{n_N}, 
\end{equation}
\begin{equation} \label{renbiasparamDef}
 b_{n_0n_1\cdots n_N} = \int \frac{d\mathbf{\Lambda}}{(2\pi)^{N+1}} e^{-\frac{1}{2}\mathbf{\Lambda}^\text{T}\mathbf{\Sigma}\mathbf{\Lambda}}
 \tilde{F}(\mathbf{\Lambda})(i\lambda)^{n_0}(i\eta_{\nabla^2\delta})^{n_1} \cdots (i\eta_{\nabla^{2N}\delta})^{n_N}, 
\end{equation}
with $\mathbf{\Lambda} = (\lambda, \bar{\eta}_{\nabla^2 \delta}, \dots, \bar{\eta}_{\nabla^{2N} \delta})$ %and similar generalization for $\mathbf{\Sigma}$. 
and $\Sigma_{ij}=\langle \nabla^{2(i-1)}\delta_R \nabla^{2(j-1)}\delta_R\rangle$. Analogous relations to Eq.~(\ref{bivRenToBar}) follow from substituting 
Eq.~(\ref{spectralbiasrel}) in Eq.~(\ref{renbiasparamDef}).
To check consistency with earlier work, we consider the case 
\begin{equation}
 b_{\nabla^{2N}\delta} \equiv b_{\underbrace{\scriptstyle{00\cdots 1}}_{N+1}}.
\end{equation}
Using the spectral bias parameters relation of Eq.~(\ref{spectralbiasrel}), we quickly find the relation between renormalized and bare bias parameters:
\begin{equation}
 b_{\nabla^{2N}\delta} = \sum_{m=0}^N (-1)^{N-m} W_{N-m} R_\Lambda^{2(N-m)} c_{\nabla^{2m}\delta},
\end{equation}
which is the same result presented in Eq.~(B14) of Ref.~\cite{Schmidt:2012ys}, with $[W_n]^\text{here}=[\frac{(-1)^n f_n}{(2n+1)!}]^\text{that work}$.

The connection between our formalism and the peak background split bias (PBS) framework 
\cite{Mo:1996cn,Schmidt:2012ys} is deeper than the above result. In the PBS formalism
the bias parameters are defined as responses of the mean abundance of tracers to small 
changes of background density and curvature. If the background density $\bar{\rho}$ is shifted by a constant
amount $ \Delta \bar{\rho} = \bar{\rho} D $, the smoothed overdensity is shifted as
\begin{equation}
 \bar{\rho} \rightarrow \bar{\rho}  + \bar{\rho} D: \qquad \delta_R \rightarrow   \delta_R + D. 
\end{equation}
On the other hand, a constant shift on the curvature $\nabla^2 \delta \rightarrow \nabla^2 \delta  + \alpha$
induces a shift on the Laplacian of the smoothed density and on the smoothed density itself 
\begin{equation}
 \nabla^2 \delta \rightarrow  \nabla^2 \delta  + \alpha: \qquad  
 \nabla^2 \delta_R  \rightarrow   \nabla^2 \delta_R  + \alpha, 
 \quad \delta_R \rightarrow   \delta_R - W_1 R_\Lambda^2 \alpha,
\end{equation}
where the last relation holds if the density field $\delta$ is evaluated at the center of the window function (see \cite{Schmidt:2012ys}),
otherwise subdominant terms should be added.  
For bivariate bias parameters we simply take the combined transformation, 
$\delta_R \rightarrow \delta_R + D - W_1 R_\Lambda^2 \alpha$, $\nabla^2 \delta \rightarrow  \nabla^2 \delta  + \alpha$. 
Since in this work we have defined $F$ through Eq.~(\ref{defF2}), the mean abundance of tracers $n_X$ at position $\vx$ is given by 
$n_X(\vx) = \langle n_X \rangle_{\vx} F$, and the PBS biases are
\begin{equation}
 b_{nm}^\text{PBS} = \left\langle \frac{\partial^{n+m}F(\delta+D- W_1 R_\Lambda^2 \alpha, \nabla^2 \delta +\alpha)}{\partial D^n \partial \alpha^m}
 \Big|_{\alpha=0,D=0} \right\rangle.
\end{equation}
We want to show that this coincides with our definition of bias in Eq.~(\ref{DefBivRenPar}). First, we have
\begin{equation}
F(\delta+D- W_1 R_\Lambda^2 \alpha, \nabla^2 \delta +\alpha) =  
\int \frac{d\Lambda}{2\pi}\frac{d\eta}{2\pi} e^{i \lambda \delta + i \bar{\eta} \nabla^2 \delta} F(\lambda,\bar{\eta})
e^{i D\lambda +  i \alpha( \bar{\eta} -W_1\lambda R_\Lambda^2)}.
\end{equation}
By taking derivatives with respect to $D$ and $\alpha$,
\begin{align} \label{pnmFpDa}
\frac{\partial^{n+m}F(\delta+D- W_1 R_\Lambda^2 \alpha, \nabla^2 \delta +\alpha)}{\partial D^n \partial \alpha^m}
 \Big|_{\alpha=0,D=0} &= 
 \int \frac{d\Lambda}{2\pi}\frac{d\eta}{2\pi} e^{i \lambda \delta + i \bar{\eta} \nabla^2 \delta} F(\lambda,\bar{\eta})
 (i\lambda)^n (i\bar{\eta} - iW_1 R_\Lambda^2)^m \nonumber\\
 &= \int \frac{d\mathbf{\Lambda}}{(2\pi)^2} e^{ i \mathbf{\Lambda} \cdot \mathbf{D} } F(\mathbf{\Lambda})
 (i\lambda)^n (i \eta)^m,
\end{align}
where in the last equality we have used $\eta = \bar{\eta} - W_1 R_\Lambda^2$, consistently with Eq.~(\ref{spectralbiasrelLap}).
Now, we assume ergodicity and Gaussianity to obtain the PBS biases, that is,  integration of Eq.~(\ref{pnmFpDa}) 
against $\int d \mathbf{D} (2\pi |\mathbf{\Sigma}|^{1/2})^{-1} e^{-\frac{1}{2} \mathbf{D}^T \mathbf{\Sigma}^{-1} \mathbf{D}}$
yields
\begin{align}
 b_{nm}^\text{PBS} &= 
%  \int \frac{d\mathbf{\Lambda}}{(2\pi)^2}  F(\mathbf{\Lambda})
%  (i\lambda)^n (i \eta)^m  \int \frac{d \mathbf{D}}{2\pi |\mathbf{\Sigma}|^{1/2}} e^{-\frac{1}{2} \mathbf{D}^T \mathbf{\Sigma}^{-1} \mathbf{D}}
%  e^{ i \mathbf{\Lambda} \cdot \mathbf{D} } \nonumber\\& =  
 \int \frac{d\mathbf{\Lambda}}{(2\pi)^2} e^{-\frac{1}{2} \mathbf{\Lambda}^T \mathbf{\Sigma} \mathbf{\Lambda}} \tilde{F}(\mathbf{\Lambda})
     (i\lambda)^n (i \eta)^m = b_{nm},
\end{align}
which shows the equivalence of our renormalized bias with the PBS biases. 

We emphasize that our results rely on the reparametrization of spectral bias parameters of Eq.~(\ref{spectralbiasrel}).
Any multivariate renormalized bias parameter is found through Eq.~(\ref{renbiasparamDef}), and to find the relation between the
$c_{n_1n_2\dots}$ and $b_{n_1n_2\dots}$ one uses Eq.~(\ref{spectralbiasrel}) and performs combinatorial algebra.

\end{subsection}

\end{section}

\begin{section}{Renormalization of contact terms}\label{Sect:4}

%The higher-derivative
%terms will similarly have stochastic counterparts, for example ε∇2δ∇2δ.

The SPT power spectrum corresponding to Eq.~(\ref{LPTPSgradD}) is
\begin{align}\label{PShdt}
 P^\text{SPT}_X(k) &\ni - 2 (1 + b_{10}) b_{01} k^2 P_L(k) 
 -2(b_{20}b_{01} + b_{10}b_{11} ) \int \Dk{p} \frac{(\vk-\vp)^2 \vk \cdot \vp}{p^2}P_L(|\vk-\vp|)P_L(p) \nonumber\\
 &\quad - 2 b_{20}b_{11} \int \Dk{p} p^2 P_L(|\vk-\vp|)P_L(p).
\end{align}
The last contribution, named here as $\bar{\mathcal{I}}^{(0,2)}(k)$,
is analogous to the function $a_{02}(k)$ that we found in Sec. \ref{Sect:2}. But, this time the UV divergence cannot be
removed by a white noise absorbing a constant $\bar{\mathcal{I}}^{(0,2)}(k=0) = \int_\vp p^2 P_L^2(p)$.
It is clear that the situation will still get worse as higher derivatives are considered.  For example, in Eq.~(\ref{Bias2orderExp}) we find the term
$b_{02}^2 (\nabla^4 \xi)^2$, leading to a SPT power spectrum contribution $\int_{|\vp|<\Lambda}  (\vk - \vp)^4 p^4 P_L(|\vk-\vp|)P_L(p) $, which scales as 
$\Lambda^{10+2n}$, with $n$ the spectral index of the linear power spectrum at small scales. 
The same divergences are present in the Eulerian treatment of bias \cite{Assassi:2014fva}, and it is well known that
these can be absorbed by the stochastic bias. 
The subject of this section is to provide a systematic procedure to remove these divergences.

Stochastic fields are, by construction, uncorrelated with long wavelength perturbations 
and among themselves at scales beyond $r>1/\Lambda$ \cite{Dekel:1998eq}. In principle, they 
contain all the nonlinear processes that we have smoothed, as much as in the EFTofLSS \cite{Baumann:2010tm}. 
In Fourier space its 2-point function is commonly written as \cite{2009JCAP...08..020M,Desjacques:2016bnm} 
 \begin{equation}\label{stochcorr}
\langle \epsilon(\vk) \epsilon(\vk')\rangle' = N_0 + N_1 k^2 + N_2 k^4 +\cdots .
\end{equation}
The departure of a white noise arises because stochasticity is not localized at a single point; 
instead it is a nonlocal process with a range of coherence 
$\sim 1/\Lambda$, and for the same reason the above description breaks down for $k \sim \Lambda$. 
The absence of odd powers of $k$ comes from the spherical symmetry of the filter.

The last term in Eq.~(\ref{PShdt}) is the Fourier transform of one of the many 
products of correlators evaluated at the same point (commonly named contact terms) 
we find in the correlation function. In full generality, we find integrals as 
\begin{equation} \label{barI}
\bar{\mathcal{I}}^{(s,t)}(k,\Lambda) \equiv (-1)^{t+s}\mathcal{F}[(\nabla^{2s}\xi(\vq))(\nabla^{2t}\xi(\vq))] 
= \int \Dk{p}  |\vk-\vp|^{2t} p^{2s} P_L(|\vk-\vp|)P_L(p),
\end{equation}
where we have written explicitly the cutoff dependence. For a scale invariant power spectrum, $P_L\propto p^n$, 
$\bar{\mathcal{I}}^{(s,t)}$ scales as $\Lambda^{2(n+s+t+1)}$. 
Our goal is to find a renormalized function $\mathcal{I}^{(s,t)}$ with two properties: first, that it is UV safe (for $n<-1/2$, for example), 
and second, that it can be written as the sum of the bare function (Eq.~(\ref{barI})) and a finite set of counterterms: 
\begin{equation} \label{ctAsNoise}
 \mathcal{I}^{(s,t)}(k)= \bar{\mathcal{I}}^{(s,t)}(k,\Lambda) + \mathcal{I}^{(s,t)}_{ct}(k,\Lambda), 
 \end{equation}
with counterterms
\begin{equation}
 \mathcal{I}^{(s,t)}_{ct}(k,\Lambda) =\mathcal{I}_0(\Lambda) 
 + \mathcal{I}_1(\Lambda) k^2 + \cdots + \mathcal{I}_m(\Lambda) k^{2m},
\end{equation}
for some integer $m$ and with $\mathcal{I}_i(\Lambda)$ depending only on the cutoff $\Lambda$. In such a way, the counterterms can be 
absorbed by stochastic bias terms.
We  first write the expansion 
\begin{equation} \label{expkP}
 |\vk-\vp|^{2t} P_L(|\vk-\vp|) =p^{2t} P_L(p) \sum_{\ell=0}^{\infty} c_\ell(\mu^\ell, \mu^{\ell-2},\dots) \frac{k^\ell}{p^\ell},
\end{equation}
where we have assumed $p^s P^{(s)}_L(p) \propto P_L(p)$, which is a good approximation for
large $p$ and holds exactly for scale invariant universes. The $c_\ell$ are polynomials of $\mu = \hat{\vk}\cdot\hat{\vp}$ with
$c_0=1$ and containing only even (odd) powers of $\mu$ if $\ell$ is even (odd).
We propose the counterterms
\begin{equation}\label{ctct}
\mathcal{I}_{ct}^{(s,t)}(\vk,\Lambda) = -\int \Dk{p} p^{2(s+t)} (P_L(p))^2   
\sum_{\ell=0}^{2(s+t)} c_\ell(\mu^\ell, \mu^{\ell-2},\dots) \frac{k^\ell}{p^\ell}.
\end{equation}
Thus our candidate for renormalized $\mathcal{I}$ is
\begin{align} \label{RI}
 \mathcal{I}^{(s,t)}(k)&= \int \Dk{p} p^{2s} P_L(p)\left(|\vk-\vp|^{2t} P_L(|\vk-\vp|) 
   - p^{2t} P_L(p) \sum_{\ell=0}^{2(s+t)} c_{\ell}(\mu^\ell, \mu^{\ell-2},\dots) \frac{k^\ell}{p^\ell} \right).  
\end{align}
By plugging in the expansion (\ref{expkP}) into the above equation we get
\begin{align} \label{RI2}
 \mathcal{I}^{(s,t)}(k)
 &= \sum_{\ell=2(s+t)+1}^{\infty} k^\ell \int \Dk{p} c_{\ell}(\mu^\ell, \mu^{\ell-2},\dots) p^{2(s+t)-\ell} (P_L(p))^2 
 =    \alpha k^{2(s+t)+2} \int \Dk{p} \frac{(P_L(p))^2}{p^2} + \cdots , 
\end{align}
where the first term in the sum of the first equality vanishes when performing the angular integral because $c_{2(s+t)+1}$ is odd in $\mu$. 
% hus
% we are left with
% \begin{align}
%  \mathcal{I}^{(s,t)}(\vk)&= \alpha k^{2(s+t)+2} \int \Dk{p} \frac{(P_L(p))^2}{p^2} + \cdots    
% \end{align}
The last equality shows that the renormalized function $\mathcal{I}$  
is UV safe for spectral index $n<-1/2$ (the rest of the terms, not shown, are even more convergent; 
here $\alpha = 2 \int_{-1}^{1}d\mu \, c_{2(s+t)+2}$ is a number.
To complete the proof that $\mathcal{I}^{(s,t)}$ is indeed the renormalized function we are searching for,
we need to show that the counterterms can be written in the form of Eq.~(\ref{ctAsNoise}). This follows immediately from Eq.~(\ref{ctct})
because contributions with $\ell$ odd vanish, and  we can write
\begin{equation}\label{ctexp}
 \mathcal{I}_{ct}^{(s,t)}(k,\Lambda) = -\sum_{i=0}^{s+t} \alpha_i k^{2i} \int \Dk{p} p^{2(s+t-i)} (P_L(p))^2.
\end{equation}

At this point one may wonder if this procedure can be continued indefinitely by cutting the sum in Eq.~(\ref{ctct}) 
at some $N>2(s+t)$ and making the renormalized function $\mathcal{I}$ as close
to zero as desired. While this is true, it would require an arbitrary 
number of stochastic bias operators, as we see below.

We can accommodate the stochastic fields in the approach we have followed by adding $\epsilon(\vq)$ as a new argument to the Lagrangian bias function $F$, 
which introduces a spectral bias $\lambda_\epsilon$ and
corresponding bias parameters $b_\epsilon, b_{\epsilon^2},\dots$. Thereafter we expand similarly to Eq.~(\ref{expdR}),
$\epsilon(\vq) \simeq \epsilon_0(\vq) + \alpha R_\Lambda^2 \nabla^2 \epsilon_0(\vq) + \cdots$, with $\alpha$ a number whose value is not important for our discussion. 
This suggests introducing a second stochastic bias operator $\nabla^2 \epsilon$ with its own 
bias spectral parameter $\eta_{\nabla^{2}\epsilon}$ and a set of bias parameters $\{b_{(\nabla^{2}\epsilon)^N}\}$. Since 
the stochastic terms are uncorrelated with long wavelength overdensities
and among themselves at large scales, we have $\langle \epsilon(\vq_1)\epsilon(\vq_2)\rangle_{|\vq|>R_{\Lambda}} = 0$ and
$\langle \epsilon(\vq_1)\delta_R(\vq_2)\rangle = 0$, and equivalent equations for $\nabla^2 \epsilon$. 
Accordingly, the covariance matrix $\mathbf{\Sigma}$ will have 
an isolated block including only zero-lag correlators of stochastic fields, 
and connected correlators of stochastic
fields will be singled out in tracer statistics.\footnote{This is not entirely true because we still have contributions, as 
$\langle \epsilon \Delta_i \rangle$, different from zero at small scales.}  Up to second order in stochastic bias expansion we found 
terms up to $(b_{(\nabla^2\epsilon)^2}\langle \nabla^2\epsilon \nabla^2\epsilon \rangle)^2$, more precisely, we get 
the SPT power spectrum for stochastic fields
\begin{align} \label{PSepsilon}
P_\epsilon (\vk) &= \int d^3q e^{-i \vk \cdot \vq} \Big( b^2_{\epsilon} \xi_\epsilon + b_{\epsilon}b_{\nabla^2\epsilon} \nabla^2\xi_\epsilon
+ b_{\nabla^2\epsilon}^2  \nabla^4\xi_\epsilon + \frac{1}{2}b^2_{\epsilon^2}\xi_\epsilon^2 +   
(b_{\epsilon^2}b_{(\nabla^2 \epsilon)^2} + b_{\epsilon}^2b_{\nabla^2 \epsilon}^2)   (\nabla^2 \xi_\epsilon)^2 +
\frac{1}{2}b^2_{(\nabla^2\epsilon)^2}(\nabla^4 \xi_\epsilon)^2 \nonumber\\
&\quad \qquad+ 2 b_{\epsilon}^2 b_{\epsilon, \nabla^2 \epsilon} \xi_\epsilon \nabla^2\xi_{\epsilon} + 
b_{\epsilon, \nabla^2 \epsilon}^2  \xi_\epsilon \nabla^4\xi_{\epsilon}
+ b_{\epsilon, \nabla^2 \epsilon}b_{(\nabla^2 \epsilon)^2}\nabla^2\xi_{\epsilon}\nabla^4\xi_{\epsilon}) \nonumber\\
&=(a_0 + a_1 k^2 + a_2 k^4) \mathcal{F}[\xi_\epsilon] + (b_0 + b_1 k^2 + b_2 k^4 + b_3 k^6 + b_4 k^8) \mathcal{F}[\xi_\epsilon^2]
\end{align}
with
\begin{equation} \label{xiepsilon}
 \xi_\epsilon(q) \equiv \langle \epsilon_0(\vq_1) \epsilon_0(\vq_1 + \vq) \rangle,
\end{equation}
and $a_0,\dots, b_4$ combinations of the stochastic bias parameters. Since $\xi_\epsilon(q)$ vanishes quickly for $q > R_\Lambda$, we expect 
that $\mathcal{F}[\xi_\epsilon]$ and $\mathcal{F}[\xi_\epsilon^2]$ depend weakly on $k$ for $k\ll \Lambda$, being effectively constants at large scales.
It is common to assign $\xi_\epsilon(q) = P^{\{0\}}_{\epsilon} \delta_\text{D}(\vq)$ \cite{Desjacques:2016bnm}, leading to  
$ \mathcal{F}[\xi_\epsilon] =  P^{\{0\}}_{\epsilon}$, but ill defined for $\mathcal{F}[\xi_\epsilon^2]$. 

The derivation of  Eq.~(\ref{PSepsilon}) was informal because our model for stochasticity is far from being rigorous.
Hence it should be considered as indicative, almost illustrative; c.f. Sec. 2.8 of \cite{Desjacques:2016bnm}. 
However, our objective was to make notice that with the introduction of two stochastic bias operators
$\epsilon$ and $\nabla^2\epsilon$ and with a second order bias expansion, we obtain powers up to $k^8$ in Eq.~(\ref{PSepsilon}). This is the equivalent to Eq.~(\ref{Bias2orderExp}) where we consider bias operators
$\delta$ and $\nabla^2\delta$  and we 
have contact terms up to $b_{02}^2 (\nabla^4\xi)^2$, which has $s+t=4$, and Eq.~(\ref{ctexp}) contains powers up to $k^8$ also. 
Therefore, to remove the UV divergences of contact terms  we consider a Lagrangian bias function 
$F(\delta,\nabla^2\delta,\epsilon,\nabla^2\epsilon)$. If we add an argument $\nabla^4\delta$, in order to preserve the same level of convergence 
we should add $\nabla^4\epsilon$ as well.

% We attach to use as less as 
% possible number of counterterms to perform the cancelations. At second order in gradient bias expansion we need only $2+2=4$, where we 
% have contact terms up to $b_{02}^2 (\nabla^4\xi)^2$, requiring
% to cut the stochastic bias expansion (Eq.~(\ref{stochcorr})) at a power $k^{8}$.

% For the case which we have followed
% in Sect. and in appendix , we only introduce local and gradient bias $\eta \nabla^2\delta$, to perform the above cancelations in all contact
% terms in Eq.~() we must add to the function
% $X$ in Eq.~() two terms $\lambda_\epsilon \epsilon + \lambda_{\nabla^2 \epsilon} \nabla^2\epsilon $. Therefore, the number of stochastic bias introduced 
% must be at least the number of higher-derivative operators added. 

% Now, since the stochastic terms are uncorrelated with long wavelenght overdensities
% and among them at large scales, we have $\langle \epsilon(\vq_1)\epsilon(\vq_2)\rangle_{|\vq|>R_{\Lambda}} = 0$ and
% $\langle \epsilon(\vq_1)\delta_R(\vq_2)\rangle = 0$, and only the terms  $\langle \epsilon(\vq_1)\epsilon(\vq_2)\rangle_{|\vq|<R_{\Lambda}}$, 
% $\langle \epsilon(\vq_1) \nabla^2 \epsilon(\vq_2)\rangle_{|\vq|<R_{\Lambda}}$, 
% $\langle \nabla^2\epsilon(\vq_1) \nabla^2 \epsilon(\vq_2)\rangle_{|\vq|<R_{\Lambda}}$ will appear in Eq.~(), but since these are localized over 
% regions smaller than the smoothing scale, it is not necessary to write them explicitly, though is good to keep in mind they are present. 

\end{section}

\begin{section}{Conclusions}\label{Sect:conclusions}

In this work we proposed a novel method of renormalization of Lagrangian bias, 
consisting of a reparametrization of a set of spectral parameters, which we define as 
the arguments of the Fourier transformed Lagrangian bias function $\tilde{F}(\lambda,\eta_{\nabla^2\delta},\cdots)$. From this 
renormalized spectra one can easily find any multivariate bias parameter as a function 
of the bare biases. Our definition for nonlocal bias is an extension of the local case introduced by 
Matsubara \cite{Matsubara:2008wx} in order to include curvature and higher-derivative operators. We noticed that 
the local biases were already renormalized in the sense that 2-points statistics of biased tracers contain only connected moments. 
However, we find the necessity of add curvature bias to remove subleading dependencies on the smoothing scale. We have restricted our discussion to
Gaussian fields and 2-point functions nonlinearly evolved by PT, but it would be attractive to generalize our results to $N$-point statistics and
non-Gaussian initial conditions.

We checked the consistency of our results by comparing to the PBS biases of \cite{Schmidt:2012ys}. In fact, our
multivariate bias parameters are shown to be equivalently obtained from the PBS argument in the case of initial 
Gaussian fields. We believe that our renormalization is simpler than previous methods because it only relies on 
one relation between bare and bias spectral parameters---Eq.~(\ref{spectralbiasrel}), which is a key result of this work.

We further developed a systematic procedure to remove UV divergences of Fourier transformed 
contact terms. Although it was known from previous works that this is doable due to stochasticity, 
to our knowledge this is the first time that an explicit method to do it is presented. Our model for stochasticity is primitive, 
but we find it well motivated, and indeed it provides the necessary contributions to the power spectrum in order 
to absorb the UV divergences coming from contact terms.

An obvious extension of this work is to additionally consider nonlinear biases such as tidal bias. We do not foresee major complications for doing so following the 
same path of Ref.\cite{Vlah:2016bcl}. Another interesting direction is to pursue similar methods to the case
of Eulerian bias renormalization; such an endeavor would require the introduction of all bias operators consistent with the 
symmetries of the fluid equations up to the desired order in PT, including operators that cannot be expressed in terms of matter overdensities \cite{Assassi:2014fva}. 

Finally, we note that similar renormalization schemes (such as that of contact terms) are required in the EFTofLSS;
hence we believe the methods developed here can find applicability in that theory.

\end{section}

\begin{acknowledgments}
I would like to thank Jorge L. Cervantes-Cota for enlightening discussions. 
I acknowledge partial support by Conacyt Fronteras Project 281 and Conacyt project 283151. 
 
\end{acknowledgments}

\appendix

\begin{section}{2nd order in curvature and local bias expansion}\label{app:BiasExp}

In this appendix we provide an equation for the LPT power spectrum with local and curvature bias up to second order in bias expansion. 
We replace Eq.~(\ref{defX}) by
\begin{equation}
 X = \lambda_1 \delta_1 + \lambda_2 \delta_2 + \eta_1 \nabla^2 \delta_1 +  \eta_2 \nabla^2 \delta_2 + \vk \cdot \Delta
\end{equation}
and define
\begin{align}
 \sigma^2_{R} = \langle (\delta_R )^2 \rangle_c, \qquad 
 \sigma^2_{\delta\nabla^2\delta} &= \langle \delta_R \nabla^2 \delta_R \rangle_c, \qquad 
 \sigma^2_{\nabla^2\delta} = \langle (\nabla^2 \delta_R)^2 \rangle_c. 
\end{align}
With this, we get
\begin{align}
-\frac{1}{2}\langle X^2 \rangle_c &=  
 -\frac{1}{2} (\lambda_1^2 + \lambda_2^2) \sigma^2_R -\frac{1}{2} (\eta_1^2 + \eta_2^2) \sigma^2_{\nabla^2\delta}
 - (\lambda_1\eta_1 + \lambda_2\eta_2) \sigma^2_{\delta \nabla^2\delta}
   - \lambda_1 \lambda_2 \xi_R - (\lambda_1 + \lambda_2) k_i U_i -\frac{1}{2}k_ik_j A_{ij}   \nonumber\\
&\quad - (\lambda_1\eta_2 + \lambda_2\eta_1) \nabla^2\xi_R + (\eta_1 + \eta_2) k_i \nabla_i \xi_R + \eta_1 \eta_2 \nabla^4 \xi_R,             
\end{align}
and
\begin{align}
-\frac{i}{6}\langle X^3 \rangle_c &= - \frac{i}{2}(\lambda_1^2 + \lambda_2^2) k_i U^{2000}_i - \frac{i}{2}(\eta_1^2 + \eta_2^2) k_i U^{0020}_i
- i \lambda_1\lambda_2 k_i U^{1100}_i - i \eta_1 \eta_2 k_i U^{0011} - i (\lambda_1 \eta_1 + \lambda_2\eta_2)k_i U^{1010} \nonumber\\
&\quad - i (\lambda_1 \eta_2 + \lambda_1\eta_2)k_i U^{1001}- \frac{i}{2}(\lambda_1 + \lambda_2)k_i k_j A_{ij}^{1000}
- \frac{i}{2}(\eta_1 + \eta_2)k_i k_j A_{ij}^{0010}.
\end{align}

The LPT power spectrum is
% \begin{align}\label{Bias2orderExp}
% &(2\pi)^3\delta_\text{D}(\vk) + P^\text{LPT}_X(k) = \int d^3 q e^{i\vk\cdot\vq}e^{-\frac{1}{2}k_ik_jA_{ij} - \frac{i}{6}k_ik_jk_kW_{ijk}} \Big[
% 1 + b_{10}^2 \xi  + 2 i b_{10} k_i U_i + 2 b_{10} b_{01}\nabla^2 \xi  - 2 i b_{01} k_i \nabla_i \xi  \nonumber\\
%  &\quad + 2 b^2_{01} \nabla^4\xi  + \frac{1}{2} b_{20}^2 \xi^2  -  (b_{10}^2 + b_{20}) k_i k_j U_i U_j +(b_{20}b_{02} + b_{11}^2) (\nabla^2\xi )^2 
%  - (b_{02} + b_{01}^2) k_ik_j \nabla_i \xi  \nabla_j \xi  + \frac{1}{2} b^2_{02} (\nabla^4 \xi )^2 \nonumber\\
%  &\quad + 2 i b_{10} b_{20} \xi  k_i U_i 
%  + 2 b_{20}b_{11}\xi   \nabla^2\xi  - 2 i b_{10} b_{11} \xi  k_i \nabla_i \xi  + 2i (b_{20}b_{01} + b_{10}b_{11}) \nabla^2\xi   k_i U_i \nonumber\\
%  &\quad + 2 (b_{11} + b_{10}b_{01}) k_i k_j \nabla_i \xi  U_j
%          +  2 b_{11}b_{01}  k_i U_j  \nabla^4 \xi  -2(b_{11}b_{01} + b_{10}b_{02}) k_i \nabla_i \xi \nabla^2\xi  \nonumber\\
%  &\quad+2 b_{11}b_{02} \nabla^2\xi  \nabla^4\xi  
%        -  2 i b_{01}b_{02} k_i \nabla_i \xi  \nabla^4\xi  + i b_{20} k_i U^{2000}_i + i b_{02} k_i U^{0020}_i
%        + i b_{10}^2 k_i U^{1100}_i + i b_{01}^2 k_i U^{0011}  \nonumber\\
% &\quad + 2 i b_{11} k_i U^{1010} + 2 i b_{10}b_{01} k_i U^{1001}- b_{10} k_i k_j A_{ij}^{1000}
%    - b_{01} k_i k_j A_{ij}^{0010} \Big]
% \end{align} 
%%%% CORRECCION %%%%%%%%%%%%%%%%%%%%
\begin{align}\label{Bias2orderExp}
&(2\pi)^3\delta_\text{D}(\vk) + P^\text{LPT}_X(k) = \int d^3 q e^{i\vk\cdot\vq}e^{-\frac{1}{2}k_ik_jA_{ij} - \frac{i}{6}k_ik_jk_kW_{ijk}} \Big[
1 + b_{10}^2 \xi  + 2 i b_{10} k_i U_i + 2 b_{10} b_{01}\nabla^2 \xi  - 2 i b_{01} k_i \nabla_i \xi  \nonumber\\
 &\quad +  b^2_{01} \nabla^4\xi  + \frac{1}{2} b_{20}^2 \xi^2  -  (b_{10}^2 + b_{20}) k_i k_j U_i U_j +(b_{20}b_{02} + b_{11}^2) (\nabla^2\xi )^2 
 - (b_{02} + b_{01}^2) k_ik_j \nabla_i \xi  \nabla_j \xi  + \frac{1}{2} b^2_{02} (\nabla^4 \xi )^2 \nonumber\\
 &\quad + 2 i b_{10} b_{20} \xi  k_i U_i 
 + 2 b_{20}b_{11}\xi   \nabla^2\xi  - 2 i b_{10} b_{11} \xi  k_i \nabla_i \xi  + b_{11}^2 \xi  \nabla^4 \xi  + 2i (b_{20}b_{01} + b_{10}b_{11}) \nabla^2\xi   k_i U_i \nonumber\\
 &\quad + 2 (b_{11} + b_{10}b_{01}) k_i k_j \nabla_i \xi  U_j +  2 i b_{11}b_{01}  k_i U_j  \nabla^4 \xi  -2i(b_{11}b_{01} + b_{10}b_{02}) k_i \nabla_i \xi \nabla^2\xi  \nonumber\\
 &\quad+2 b_{11}b_{02} \nabla^2\xi  \nabla^4\xi  
       -  2 i b_{01}b_{02} k_i \nabla_i \xi  \nabla^4\xi  + i b_{20} k_i U^{2000}_i + i b_{02} k_i U^{0020}_i
+ i b_{10}^2 k_i U^{1100}_i + i b_{01}^2 k_i U^{0011}  \nonumber\\
&\quad + 2 i b_{11} k_i U^{1010} + 2 i b_{10}b_{01} k_i U^{1001}- b_{10} k_i k_j A_{ij}^{1000}
- b_{01} k_i k_j A_{ij}^{0010} \Big]
\end{align}
where we defined, as generalizations of Eq.(\ref{defqfuncts}), 
\begin{equation}
 U^{pqrs}_i(\vq) = \langle \delta_1^p \delta_2^q (\nabla^2 \delta_1)^r (\nabla^2 \delta_2)^s \Delta_i \rangle_c, \qquad
 A_{ij}^{pqrs}= \langle \delta_1^p \delta_2^q (\nabla^2 \delta_1)^r (\nabla^2 \delta_2)^s \Delta_i \Delta_j \rangle_c.
\end{equation}
The following identities are valid
\begin{align}
& U^{2000}= U^{0200}=U^{20}, \qquad U^{1100}=U^{11}, \qquad U^{0020} = U^{0002} \\
& U^{1010}=U^{0101},\qquad U^{1001} = U^{0110} \\
& A_{ij}^{1000} = A_{ij}^{0100} = A_{ij}^{10}, \qquad A_{ij}^{0010} = A_{ij}^{0001}.
\end{align}

We note that we have used the renormalized bias to write Eq.~(\ref{Bias2orderExp}) instead of the biases $c_{nm}$, and get rid
of the label $R$ in the linear correlation function. That is, we have assumed that all dependences on $R_\Lambda$ were removed
for $q>R_{\Lambda}$. We also omitted to write the stochastic field contributions.

\end{section}

 \bibliographystyle{JHEP}  % Use the "unsrtnat" BibTeX style for formatting the Bibliography
 \bibliography{biasbib.bib}

 \end{document}